\documentclass[aps,twocolumn,showpacs,floats,prl]{revtex4}
\usepackage{graphicx}
\usepackage{dcolumn}
\usepackage{bm}
\usepackage{color}
\usepackage{subfigure}
\usepackage{amsmath}

\begin{document}

\author{L. Parisi$^1$, G. E. Astrakharchik$^2$, and S. Giorgini$^1$}

\affiliation{
$^1$ Dipartimento di Fisica, Universit\`a di Trento and CNR-INO BEC Center, I-38050 Povo, Trento, Italy \\
$^2$ Departament de F\'{i}sica i Enginyeria Nuclear, Universitat Polit\`ecnica de Catalunya, Campus Nord B4-B5, E-08034, Barcelona, Spain
}

\title{Spin dynamics and Andreev-Bashkin effect in mixtures of one-dimensional Bose gases} 

\begin{abstract} 
We investigate the propagation of spin waves in two-component mixtures of one-dimensional Bose gases interacting through repulsive contact potentials. By using quantum Monte Carlo methods we calculate static ground-state properties, such as the spin susceptibility and the spin structure factor, as a function of the coupling strengths and we determine the critical parameters for phase separation. In homogeneous mixtures, results of the velocity of spin waves and of its softening close to the critical point of phase separation are obtained by means of hydrodynamic theory and a sum-rule approach. We quantify the non-dissipative drag effect, resulting from the Andreev-Bashkin current-current interaction between the two components of the gas, and we show that in the regime of strong coupling it causes a significant suppression of the spin-wave velocity. \end{abstract}
\maketitle 

The problem of dissipationless spin transport is a widely studied topic in condensed matter physics with important applications to electron-hole superfluidity, superfluid $^3$He and spintronic devices~\cite{Sonin}. Ultracold gases, with the possibility they offer to realize  quantum degenerate mixtures, open new interesting perspectives for the investigation of spin dynamics. Spin diffusion in a strongly interacting two-component Fermi gas has been observed and characterized in a series of recent experiments~\cite{Sommer11, Bardon14, Valtolina17}, whereas the existence of spin supercurrents in Bose mixtures has been demonstrated both at very low temperatures~\cite{Modugno02, Mertes07, Nicklas11, Zhang12, Egorov13, TN-exp1} and in the presence of a large thermal component~\cite{TN-exp2}. In this respect one-dimensional (1D) mixtures are particularly interesting for several reasons: i) the low-energy dynamics is universal and described by the Luttinger liquid model~\cite{Voit94}; ii) spin and charge degrees of freedom are expected to be completely decoupled at low energy~\cite{Recati03-1, Recati03-2}; and, finally, iii) regimes of strong interactions can be achieved in long-lived samples~\cite{Stoferle04, Paredes04, Weiss05, Haller09}.

The undamped propagation of spin waves is an important signature of spin superfluidity and an unbiased determination of the spin-sound velocity is a crucial element to understand the dynamics of two-component Bose mixtures at ultralow temperatures. Notably, for such mixtures, the propagation of sound in the spin channel depends not only on the static magnetic susceptibility, but also on a purely dynamic quantity known as the Andreev-Bashkin non-dissipative drag~\cite{AndreevBashkin}. This intriguing effect, never observed so far, involves two coupled superfluids and entails that a superflow in one component can induce a supercurrent in the second component which is dragged without energy dissipation. In its original form, the Andreev-Bashkin effect was discussed in connection with possible superfluid mixtures of $^3$He in $^4$He. However, due to the limited solubility of the two isotopes~\cite{Helium}, such superfluid mixtures have never been realized. In the context of ultracold atoms the Andreev-Bashkin effect was studied in the continuum using a perturbative approach based on the Bogoliubov theory~\cite{FilShevchenko} as well as in lattice systems~\cite{Kaurov05}. More recently, its consequences on the propagation of spin waves were analyzed using the hydrodynamic theory~\cite{Nespolo17}. 

In the present work we investigate spin dynamics and the effect of the Andreev-Bashkin superfluid drag in 1D repulsive mixtures of Bose gases. To this aim we use quantum Monte-Carlo (QMC) methods first to establish the critical condition for the miscibility of the two gases, which is a preliminary requisite to investigate the regime of homogeneous mixtures. Second we calculate the entrainment effect from the coupled superfluid response and the spin-wave velocity by means of hydrodynamic theory and of a sum-rule approach. On the basis of simulations performed by varying both the intra-species and the inter-species coupling strength, we find that the superfluid drag can be large if the inter-species coupling is strong, and it contributes to the softening of spin waves on approaching the critical point of phase separation.

We consider Bose-Bose mixtures in a 1D geometry described by the following Hamiltonian
\begin{eqnarray}
H&=&-\frac{\hbar^2}{2m}\sum_{i=1}^{N_a} \frac{\partial^2}{\partial x_i^2}+g\sum_{i<j}\delta(x_i-x_j)
\label{Hamiltonian}
\\
&-& \frac{\hbar^2}{2m}\sum_{\alpha=1}^{N_b}\frac{\partial^2}{\partial x_\alpha^2}+g\sum_{\alpha<\beta}\delta(x_\alpha-x_\beta)+\tilde{g}\sum_{i,\alpha}\delta(x_i-x_\alpha) \;,
\nonumber
\end{eqnarray}
which includes, in addition to the kinetic energy terms of the two components with $N_a$ and $N_b$ particles, equal intra-species interactions modeled by the contact coupling constant $g>0$ and a contact inter-species repulsive potential of strength $\tilde{g}>0$. Here $x_i$ with $i=1,\dots,N_a$ and $x_\alpha$ with $\alpha=1,\dots,N_b$ denote, respectively, the positions of particles belonging to component $a$ and $b$ of the mixture. We also consider mass balanced mixtures, $m$ being the mass of particles of both components. In the absence of inter-species interactions, the above Hamiltonian for each component separately yields the well-known Lieb-Liniger (LL) model~\cite{LiebLiniger}, which can be solved exactly via Bethe ansatz for any value of the coupling constant $g$. In particular, for very strong repulsion ($g\to\infty$) corresponding to the so-called Tonks-Girardeau (TG) regime, the LL model describes a gas of impenetrable bosons which is equivalent to a gas of non-interacting spinless fermions~\cite{Girardeau}. The full Hamiltonian~(\ref{Hamiltonian}) consists of two LL gases, with the same interaction strength $g$, coupled via a contact repulsive force. One should point out that this full Hamiltonian also admits exact solutions, but only when it enjoys SU(2) symmetry, {\it i.e.} if $\tilde{g}=g$ or when both components are in the TG regime ($g=\infty$). In the first case the ground state is ferromagnetic~\cite{Li03, Yang03} and the equation of state can be calculated using the LL model of a single-component gas. In the latter case, the system corresponds to a mixture of interacting Fermi gases and the solution is provided by the Yang-Gaudin model~\cite{Yang, Gaudin} yielding a paramagnetic ground state for any value of the repulsive coupling $\tilde{g}$. In all other cases, for which the Bethe ansatz approach is no longer applicable, only numerical solutions are available by means, for example, of QMC methods. 

A population balanced system, where $N_a=N_b=N/2$, can be fully characterized in the thermodynamic limit by the following two dimensionless parameters
\begin{equation}
\gamma=\frac{gm}{\hbar^2n} \;\;\;\;\;\; \eta=\frac{\tilde{g}m}{\hbar^2n} \;.
\label{parameters}
\end{equation}
These are fixed by the values of interaction strength and by the total density $n=n_a+n_b$ of the gas, where $n_a=N_a/L$ and $n_b=N_b/L$ are the densities of the two components in terms of the size $L$ of the 1D box. In unbalanced configurations, an additional parameter is needed to describe the polarization: $P=(n_a-n_b)/n$. In Refs.~\cite{Parisi,Grusdt17} the ground-state energy of the Hamiltonian~(\ref{Hamiltonian}) was calculated in the extreme case of one impurity immersed in a LL gas ($N_b=1$). Here we make use of a similar diffusion Monte Carlo method extended to any configuration $N_b\leq N_a$ with periodic boundary conditions. This technique provides exact results for the ground-state energy $E(\gamma,\eta,P)$ of the mixture within statistical uncertainty~\cite{DMC}. More in details, simulations utilize a guiding wave function used for importance sampling and to encode the contact boundary conditions imposed by the $\delta$-function potentials in the Hamiltonian~(\ref{Hamiltonian}). The guiding wave function is constructed as a product of pair-wise correlation terms which, at short interparticle distance, reproduce the exact solution of the two-body problem with the contact potential, as well as many-body correlations typical of Luttinger liquids at longer distances~\cite{SuppMat1}. In this way, unbiased calculations of properties of the interacting gas are actually carried out by simulating free particles subject to proper boundary conditions~\cite{Astra03, Parisi}. The relevance of finite-size effects is estimated by repeating the simulations with increasing numbers of particles (typically ranging from $N=50$ to $N=200$), thereby ensuring a well-controlled approach to the thermodynamic limit. 
 
\begin{figure*}
\scalebox{1.0}{\includegraphics[width=0.5\textwidth,height=6.0cm]{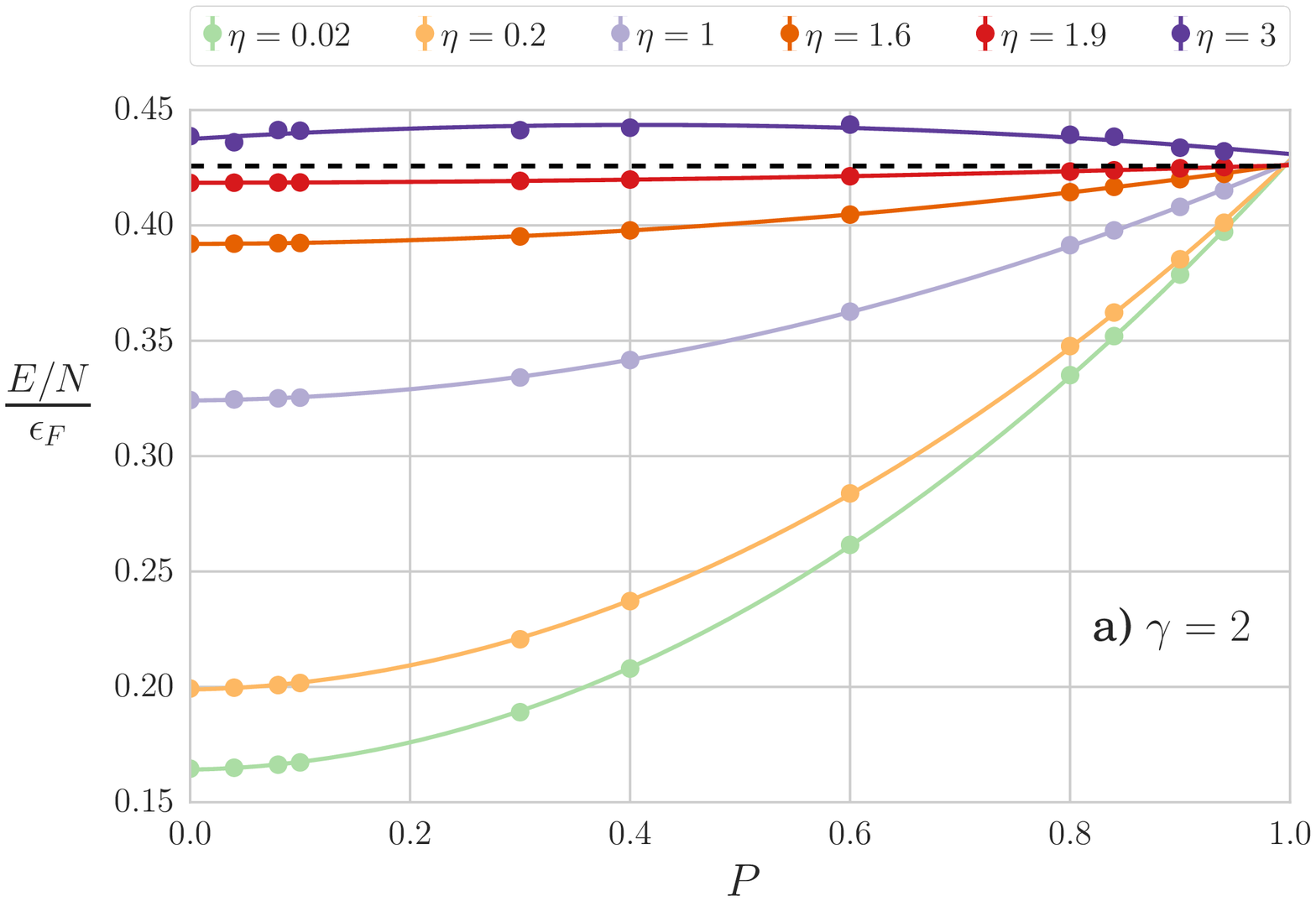}\includegraphics[width=0.5\textwidth,height=6.0cm]{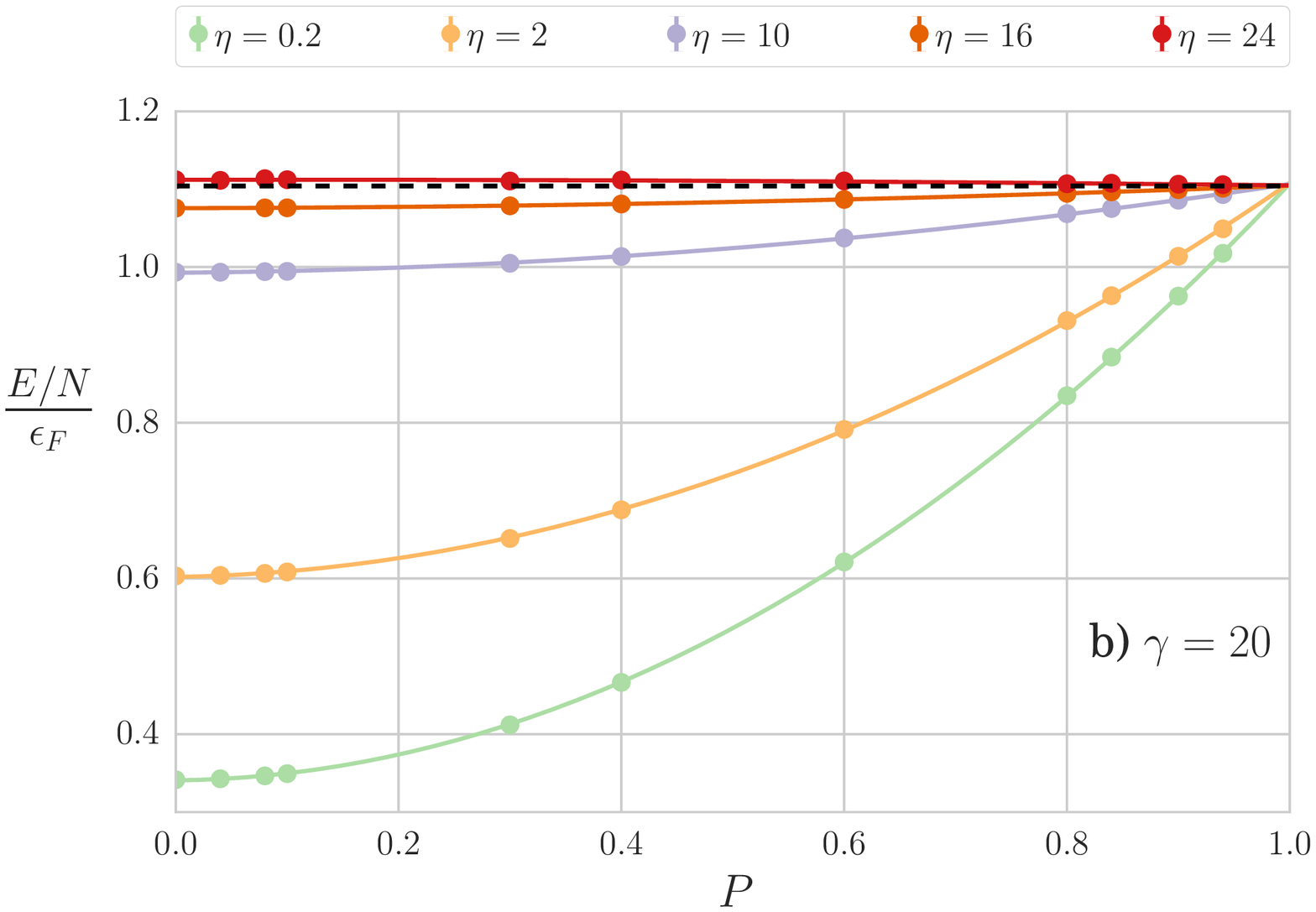}}
\caption{(color online) Energy per particle of the homogeneous mixture as a function of the polarization $P$ for different values of the coupling strengths $\eta$. Panel (a) refers to $\gamma=2$ and panel (b) to $\gamma=20$. Energies are shown in units of $\epsilon_F=\frac{\hbar^2\pi^2n^2}{8m}$ corresponding to the Fermi energy of the mixture when $\gamma=\infty$. The solid lines are best fits quadratic in $P$ and the dashed horizontal lines indicate the energy of the fully polarized ($P=1$) states. Statistical error bars are smaller than the symbol size.}
\label{fig1}
\end{figure*}

\begin{figure}
\begin{center}
\includegraphics[width=8.5cm]{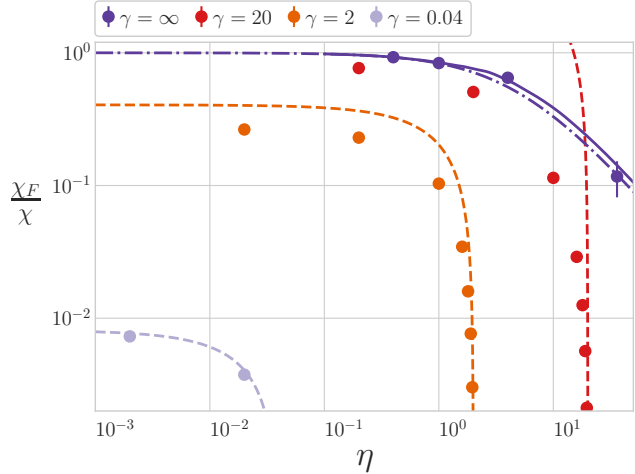}
\caption{(color online). Inverse susceptibility $1/\chi$ as a function of $\eta$ for various values of $\gamma$ ranging from the weak-coupling regime ($\gamma=0.04$) to the TG limit ($\gamma=\infty$). Here $\chi_F=\frac{4m}{\hbar^2\pi^2n}$ is the susceptibility of the non-interacting mixture when $\gamma=\infty$. Dashed lines correspond to the mean-field result $1/\chi=(g-\tilde{g})/2$, whereas the dash-dotted line to the perturbation expansion $\frac{\chi_F}{\chi}=1-\frac{2\eta}{\pi^2}$ holding in the TG limit. The solid line is obtained from the exact solution of the Yang-Gaudin model at finite polarization.}
\label{fig2}
\end{center}
\end{figure}

{\it Phase Separation:} The first question we address concerns the condition of miscibility of the mixture at $T=0$ and of its eventual phase separation. This latter is signalled by the divergence of the magnetic susceptibility $\chi$, whose inverse is related to the curvature of the energy increase as the system is polarized away from the $P=0$ balanced configuration: $\frac{1}{\chi}=\frac{\partial^2 E/L}{n^2\partial P^2}$. In the weak-coupling regime, corresponding to $\gamma\ll1$ and $\eta\ll1$, one can use the mean-field theory yielding the analytical result $\frac{1}{\chi}=\frac{g-\tilde{g}}{2}$~\cite{PitaevskiiStringari16}. Based on this approach the mixture is miscible for $\tilde{g}<g$ and phase separation occurs as soon as $\tilde{g}>g$. Furthermore, in the Yang-Gaudin model where both components are in the fermionic TG limit, the homogenous mixture is known to be stable for any value of the inter-species coupling strength $\tilde{g}$. A question worth addressing concerns the determination of the critical parameter for phase separation in the regime of intermediate values of the coupling strength $\gamma$. To this purpose we calculate the ground-state energy for fixed values of $\gamma$ and $\eta$ with varying polarization $P$. The characteristic dependencies, obtained for $\gamma=2$ and $\gamma=20$, are shown in Fig.~\ref{fig1}. We find that the energy of the $P=0$ state is lower than the one of the fully polarized ($P=1$) state provided that $\eta<\gamma$. For $\eta$ slightly larger than $\gamma$ the energy lies above the $P=1$ threshold signalling the instability against the formation of two fully polarized domains~\cite{note1}. From the equation of state as a function of the polarization $P$ we extract the inverse magnetic susceptibility $1/\chi$ which we report in Fig.~\ref{fig2} for various values of $\gamma$. We see that for $\gamma=0.04$ the results of $1/\chi$ are well reproduced by the mean-field prediction whereas, for larger values of $\gamma$, deviations are visible away from the critical point. Close to the point of phase separation, however, we notice that the susceptibility of both $\gamma=2$ and $\gamma=20$ is well described by the linear dependence $1/\chi\propto (\gamma-\eta)$ of the mean-field prediction. Finally, for $\gamma=\infty$, our results reproduce the $\eta$-dependence of $1/\chi$ obtained from the exact solution~\cite{Shiba72, Batchelor06, Guan12} of the Yang-Gaudin model at finite polarization~\cite{note2}. From this analysis we conclude that, for the reported values of $\gamma<\infty$, the critical parameter for phase separation is $\eta=\gamma$. At this value of the inter-species interaction strength the system jumps from being paramagnetic with $P=0$ to fully ferromagnetic. These results are consistent with the known findings of the Yang-Gaudin model ($\gamma=\infty$) where phase separation never occurs~\cite{Yang}, and of the SU(2) symmetric case ($\gamma=\eta$) where the stable phase is ferromagnetic~\cite{Li03, Yang03}.

\begin{figure}
\begin{center}
\includegraphics[width=9.0cm]{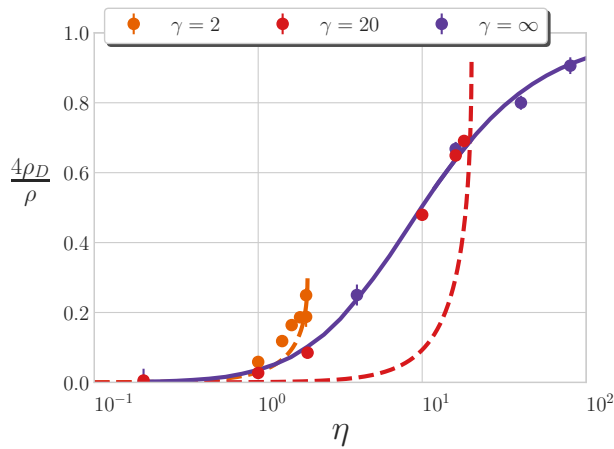}
\caption{(color online). Superfluid drag as a function of $\eta$ for different values of $\gamma$. Dashed lines correspond to the weak-coupling result (\ref{rhoD}) for $\gamma=2$ and $\gamma=20$. The solid line refers to $\gamma=\infty$ and is obtained by inverting Eq.~(\ref{spinwave1}) with both $\chi$ and $v_s$ from the exact solution of the Yang-Gaudin model.}
\label{fig3}
\end{center}
\end{figure}

{\it Superfluid drag:} In a mixture of two superfluids with mass density $\rho_1$ and $\rho_2$, the energy change per unit volume due to finite superfluid velocities ${\bf v}_1$ and ${\bf v}_2$ is given by: $\delta E=[(\rho_1-\rho_D)v_1^2+(\rho_2-\rho_D)v_2^2+2\rho_D{\bf v}_1\cdot{\bf v}_2]/2$. The quantity $\rho_D$ accounts for the coupling between the superfluids and gives rise to a drag in the superfluid current density of each component: ${\bf j}_{1,2}=\partial\delta E/\partial{\bf v}_{1,2}=(\rho_{1,2}-\rho_D){\bf v}_{1,2}+\rho_D{\bf v}_{2,1}$, known as the Andreev-Bashkin effect~\cite{AndreevBashkin}. For 1D mixtures with $\rho_1=\rho_2=\rho/2$, one can calculate $\rho_D$ to lowest order in $\tilde{g}$ by using the Bogoliubov approach of Ref.~\cite{FilShevchenko} which yields the result
\begin{equation}
\frac{\rho_D}{\rho}\simeq\frac{4\eta^2}{3\pi}\frac{1}{\left(\sqrt{2(\gamma+\eta)}+\sqrt{2(\gamma-\eta)}\right)^3}\;.
\label{rhoD}
\end{equation}
This shows that the drag effect is quadratic in the inter-species coupling $\eta$ and is maximum at the critical point $\eta=\gamma$ where it takes the value $\frac{\rho_D}{\rho}=\frac{\sqrt{\gamma}}{6\pi}$. For arbitrary coupling strengths, we calculate $\rho_D$ by means of the exact relation
\begin{equation}
4\frac{\rho_D}{\rho}=1-\lim_{\tau\to\infty}\frac{\langle\left(W_a(\tau)-W_b(\tau)\right)^2\rangle}{4ND\tau} \;,
\label{windingnumber}
\end{equation}
based on the paired superfluid response of the two components which in QMC simulations is provided by the statistics of winding numbers~\cite{Nespolo17}. Here, $D=\hbar^2/2m$ is the diffusion constant of a free particle in imaginary time $\tau=it/\hbar$, whereas the winding number $W_a(\tau)=\sum_{i=1}^{N_a}\int_0^\tau d\tau^\prime\frac{dx_i(\tau^\prime)}{d\tau^\prime}$ of the first component and, analogously, $W_b(\tau)$ of the second component are obtained by integrating the corresponding particle trajectories. In the absence of inter-species coupling, the winding numbers $W_a$ and $W_b$ are independent and, being normalized as $N_{a(b)}=\lim_{\tau\to\infty}\frac{\langle W_{a(b)}^2(\tau)\rangle}{4D\tau}$, result in $\rho_D=0$. In the opposite case of fully paired motion of the two components, the relative winding number $(W_a-W_b)$ vanishes and the drag takes its maximum value $4\rho_D/\rho=1$. In Fig.~\ref{fig3} we report the results of $\rho_D$ calculated as a function of $\eta$ for different values of the interaction parameter $\gamma$. We find that $4\rho_D/\rho$ approaches unity in the simultaneous limit $\gamma=\infty$ and $\eta=\infty$. However, already for $\gamma=20$, $\rho_D$ reaches $\sim0.7$ of its maximum value in the vicinity of the critical point $\eta=\gamma$.

{\it Velocity of spin waves:} Within the mean-field approach the long-wavelength elementary excitations in the spin channel consist of waves propagating with the velocity $v_s=\sqrt{\frac{n(g-\tilde{g})}{2m}}$~\cite{PitaevskiiStringari16}, such that $v_s^2=\frac{\rho}{m^2\chi}$ in terms of the magnetic susceptibility and of the mass density $\rho=mn$. This result, however, holds only if one neglects the drag effect exerted by one component as it moves with respect to the other. More generally, the hydrodynamic model accounting for the Andreev-Bashkin effect yields the result~\cite{Nespolo17} 
\begin{equation}
v_s^2=\frac{\rho-4\rho_D}{m^2\chi} \;,
\label{spinwave1}
\end{equation} 
which involves the superfluid drag density $\rho_D$.  

\begin{figure}
\begin{center}
\includegraphics[width=9.0cm]{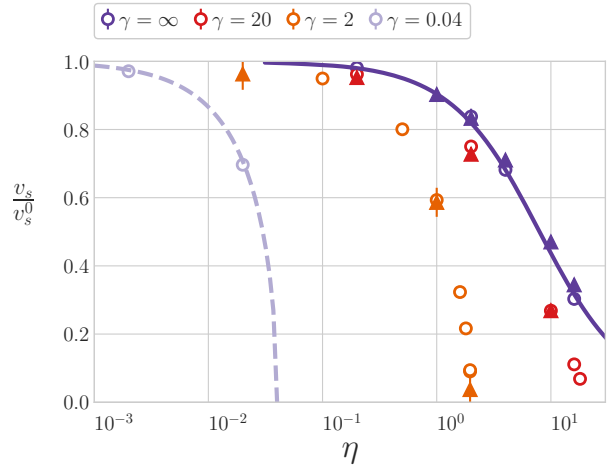}
\caption{(color online). Spin-wave velocity $v_s$ as a function of $\eta$ for different values of $\gamma$. The units are provided by the spin-wave velocity in the absence of inter-species interactions, $v_s^0=\sqrt{\rho/m^2\chi_0}$. Open symbols refer to Eq.~(\ref{spinwave1}) and solid symbols to Eq.~(\ref{spinwave2}). The dashed line corresponds to the mean-field prediction $v_s=\sqrt{n(g-\tilde{g})/2m}$ and the solid line to the exact solution in the Yang-Gaudin model~\cite{Batchelor06}.}
\label{fig4}
\end{center}
\end{figure}

In order to determine the spin-wave velocity $v_s$, we follow two independent approaches. The first is based on the hydrodynamic theory of superfluids entailed by Eq.~(\ref{spinwave1}), where we determine $v_s$ from the knowledge of the magnetic susceptibility $\chi$ and of the superfluid drag $\rho_D$ calculated above. The second, instead, is based on linear response theory and is conveniently discussed in terms of the frequency-weighted moments $m_k=\int d\omega\;(\hbar\omega)^k S_s({\bf q},\omega)$ of the spin-dependent dynamic structure factor $S_s({\bf q},\omega)$~\cite{SuppMat2}. In particular, for the following two moments one finds
\begin{eqnarray}
m_{-1}&=&N\frac{\chi_s(q)}{2n}\underset{q\to0}{\rightarrow} N\frac{\chi}{2n} \; ,
\label{m-1}\\
m_0&=&S_s(q) \;.
\label{m0}
\end{eqnarray}
Here, Eq.~(\ref{m-1}) is the susceptibility sum rule involving the static spin-spin response function $\chi_s(q)$, which reduces to $\chi$ in the long-wavelength limit, and Eq.~(\ref{m0}) defines the static spin-spin structure factor. One can show that in the $q\to0$ limit both the $m_{-1}$ and $m_0$ sum rules are exhausted by the spin-wave excitation with energy $\epsilon_s(q)=v_s\hbar q$, because multi-mode excitations contribute to the two sum rules with higher powers of $q$~\cite{PinesNozieres, DalfovoStringari, note3, SuppMat3}. From this analysis it follows that the energy of the low-lying spin-wave excitations can be obtained from the ratio of sum rules calculated in the $q\to0$ limit 
\begin{equation}
\hbar qv_s=\lim_{q\to0}\frac{m_0}{m_{-1}}\;.
\label{spinwave2}
\end{equation}
A direct calculation of the static spin-spin structure factor $S_s(q)$ allows one to extract the coefficient of its low-$q$ linear dependence $S_s(q)/N=\frac{v_s\chi}{2n}\hbar q$. Once divided by the magnetic susceptibility $\chi$ obtained above, this result gives the spin velocity $v_s$. 

Both the hydrodynamic and the more microscopic estimate of $v_s$ are shown in Fig.~\ref{fig4} and are found to agree within statistical errors providing a very strong evidence of the reliability of our results. When $\gamma=\infty$ we also find agreement with the exact result of $v_s$ from the Yang-Gaudin model~\cite{Coll74, Batchelor06}. By increasing the inter-species interaction strength, the spin-wave velocity decreases due to the combined effect of the susceptibility, which rises from the non-interacting value $\chi_0$, and of the drag density $\rho_D$ until it vanishes at the critical point of phase separation. For example, for $\eta=10$ and $\gamma=20$ the spin-wave velocity is reduced to $v_s\simeq0.3v_s^0$ of the value corresponding to $\eta=0$ (see Fig.~\ref{fig4}). The reduction is caused by an approximate five-fold increase of $\chi/\chi_0$ shown in Fig.~\ref{fig2} and by an additional factor of about 0.5 deriving from the term $1-4\rho_D/\rho$ shown in Fig.~\ref{fig3}.

In conclusion, we provide exact predictions for the velocity of spin waves in repulsive 1D Bose mixtures. These results show the strong effect of the Andreev-Bashkin superfluid drag, which could be experimentally observed by means of independent measurements of the spin-wave velocity and of the magnetic susceptibility.

{\it Acknowledgements:} We gratefully acknowledge useful discussions with A. Recati and X. W. Guan. This work was supported by the European Union's Horizon 2020 research and innovation programme under grant agreement No. 641122 ``QUIC", by Provincia Autonoma di Trento and by the grant FIS2014-56257-C2-1-P of the MICINN (Spain). G.E.A. thankfully acknowledges the computer resources at MareNostrum and the technical support provided by Barcelona Supercomputing Center (RES-FI-2017-3-0023).


\begin{thebibliography}{99}
\bibitem{note1} The instability is triggered by long-wavelength fluctuations of the magnetization density which are inhibited in our finite-size simulations. This allows us for a reliable determination of the equation of state in the metastable regime at least when $\tilde{g}$ is only slightly larger than $g$.
\bibitem{note2} We solve the coupled integral equations of the Yang-Gaudin model for polarizations $P\gtrsim10^{-3}$ and we thereby determine the susceptibility $\chi$ at small but still finite polarization. Due to logarithmic singularities~\cite{Schlottmann97, Maslov15} the calculation of $\chi$ at zero polarization would require much smaller values of $P$ which can not be reproduced by our numerical methods and are most likely experimentally irrelevant. We notice that this zero-polarization susceptibility satisfies the exact relation~\cite{Coll74} $\chi v_s=\frac{2}{\hbar\pi}$ in terms of the spin velocity $v_s$ and differs from the one shown in Fig.~\ref{fig2}.
\bibitem{note3} We notice that the energy-weighted moment (f-sum rule) $m_1=N\frac{\hbar^2q^2}{2m}$ is not exhausted at low momenta by the spin-wave excitation. In fact, its contribution to the sum rule is suppressed by the drag effect and is given by $(m_1)_{\text{sw}}=m_1(1-\frac{4\rho_D}{\rho})$.
\end{thebibliography}
\end{document}


\title{Spin dynamics and Andreev-Bashkin effect in mixtures of one-dimensional Bose gases: Supplemental Material}
\author{L. Parisi$^1$, G. E. Astrakharchik$^2$ and S. Giorgini$^1$}
\affiliation{
$^1$ Dipartimento di Fisica, Universit\`a di Trento and CNR-INO BEC Center, I-38050 Povo, Trento, Italy \\
$^2$ Departament de F\'{i}sica i Enginyeria Nuclear, Universitat Polit\`ecnica de Catalunya, Campus Nord B4-B5, E-08034, Barcelona, Spain
 }

\maketitle

\section{Choice of the guiding wave function}

The guiding wave function of the positions $x_i$ and $x_\alpha$ of particles respectively from the component $a$ and $b$ is chosen as
\begin{equation}
\psi_T(x_1,\dots,x_{N_a};x_1,\dots,x_{N_b})=\prod_{i<j}f(x_i-x_j)\prod_{\alpha<\beta}f(x_\alpha-x_\beta) \prod_{i,\alpha}h(x_i-x_\alpha) \;,
\label{trial}
\end{equation}     
where the functions $f(x)$ and $h(x)$ correspond, respectively, to intra-species and inter-species correlation terms. Both functions are built from the exact solution of the two-body problem with the contact potential up to a matching point $X_m$ as well as many-body correlations typical of Luttinger liquids. More specifically, we write  $f(x)=\sin(k|x|+\varphi(k))$ if $|x|<X_m$ and $f(x)=\sin^\beta(\pi |x|/L)$ if $X_m<|x|<L/2$. The parameters $k$ and $\beta$ are fixed by the continuity condition of the function $f(x)$ and its first derivative at the matching point $X_m$. For $|x|>X_m$ the Jastrow function takes into account long-range correlations due to phonon excitations~\cite{Reatto67} and $f^\prime(x)=0$ at $|x|=L/2$ in compliance with the periodic boundary conditions of the system. The phase shift results from the Bethe-Peierls contact condition imposed by the interatomic potential and is given by  $\varphi(k)=\arctan\frac{2k\hbar^2}{mg}$. The definition of the inter-species correlation function $h(x)$ is the same as the one above, the only difference being that in this case the phase shift is determined by the coupling constant $\tilde{g}$. The matching points $X_m<L/2$ for both correlation terms are optimized by minimizing the variational energy obtained from $\langle\psi_T|H|\psi_T\rangle$.

\section{Spin-dependent dynamic structure factor}
The spin-dependent dynamic structure factor of a many-body two-component mixture contains a wealth of information about the nature and energy spectrum of the excitations coupled to magnetic fluctuations. At zero temperature the spin-dependent dynamic structure factor is defined as the Fourier transform of the spin-spin correlation function evaluated on the ground state $|\Psi_0\rangle$:
\begin{equation}
S_s({\bf q},\omega)=\frac{1}{2\pi}\int_{-\infty}^{+\infty}dt\; e^{i\omega t} \frac{\langle\Psi_0|\rho_{-{\bf q}}^s(t)\rho_{\bf q}^s|\Psi_0\rangle}{\langle\Psi_0|\Psi_0\rangle} \;.
\label{Sqomega1}
\end{equation}
Here $\rho_{\bf q}^s=\sum_{i=1}^{N/2}e^{-iqx_i}-\sum_{\alpha=1}^{N/2}e^{-iqx_\alpha}$, with $x_i$ and $x_\alpha$ being, respectively, the coordinates of the $N_a$ and $N_b$ particles in the population balanced system, is the operator corresponding to a magnetic fluctuation with wave vector ${\bf q}$. Furthermore, $\rho_{\bf q}^s(t)=e^{iHt/\hbar}\rho_{\bf q}^se^{-iHt/\hbar}$ is the same operator following a time evolution with the Hamiltonian $H$. By introducing the complete set of energy eigenstates $|\Psi_n\rangle$, the definition of $S_s({\bf q },\omega)$ can be equivalently expressed as the positive definite sum of terms
\begin{equation}
S_s({\bf q},\omega)=\sum_{n\ge0}\delta\left(\omega-\frac{E_n-E_0}{\hbar}\right)\frac{|\langle\Psi_n|\rho_{\bf q}^s|\Psi_0\rangle|^2}{\langle\Psi_0|\Psi_0\rangle} \;,
\label{Sqomega2}
\end{equation}
involving all excited states with excitation energy $E_n-E_0$ from the ground state which are not othogonal to the magnetic perturbation $\rho_{\bf q}^s|\Psi_0\rangle$. Important relations involving the dynamic spin structure factor are provided by its energy-weighted moments $m_k=\int d\omega\;(\hbar\omega)^k S_s({\bf q},\omega)$. In particular, the inverse energy-weighted moment 
\begin{equation}
m_{-1}=\int_0^\infty d\omega \; \frac{S_s({\bf q},\omega)}{\hbar\omega}=N\frac{\chi_s(q)}{2n} \;,
\label{Sqomega3}
\end{equation}
defines the magnetic response function $\chi_s(q)$. In the long-wavelength limit this response function gives the magnetic susceptibility: $\lim_{{\bf q}\to0}\chi_s(q)=\chi$ and Eq.~(\ref{Sqomega3}) is also known as the susceptibility sum rule. The zeroth moment 
\begin{equation}
m_0=\int_0^\infty d\omega\; S_s({\bf q},\omega)=\frac{\langle\Psi_0|\rho_{-{\bf q}}^s\rho_{\bf q}^s|\Psi_0\rangle}{\langle\Psi_0|\Psi_0\rangle}=S_s({\bf q}) \;,
\label{Sqomega4}
\end{equation}
defines the static magnetic structure factor $S_s({\bf q})$, which is the Fourier transform of the magnetic pair-correlation function. The energy-weighted moment
\begin{equation}
m_1=\int_0^\infty d\omega \;\hbar\omega\; S_s({\bf q},\omega)=\frac{1}{2}\langle\Psi_0|[\rho_{-{\bf q}}^s,[H,\rho_{\bf q}^s]]|\Psi_0\rangle=N\frac{\hbar^2q^2}{2m} \;,
\label{Sqomega5}
\end{equation}
is also known as the $f-$sum rule in the spin channel. See Ref.~\cite{Hohenberg74} for a discussion of the above three sum rules. Another relevant sum rule is the cubic energy-weighted moment~\cite{DalfovoStringari89}
\begin{equation}
m_3=\int_0^\infty d\omega \;(\hbar\omega)^3\; S_s({\bf q},\omega)=\frac{1}{2}\langle\Psi_0|[[\rho_{-{\bf q}}^s,H],[H,[H,\rho_{\bf q}^s]]]|\Psi_0\rangle \;.
\label{Sqomega6}
\end{equation}
By calculating the commutators with the 1D Hamiltonian of the Bose-Bose mixture one finds\cite{DalfovoStringari89}
\begin{eqnarray}
m_3&=&N\left[\left(\frac{\hbar^2q^2}{2m}\right)^3+\left(\frac{\hbar^2q^2}{2m}\right)^2\frac{\langle\Psi_0|T|\Psi_0\rangle}{N}+\frac{\hbar^4q^2n}{4m^2}  \right. 
\nonumber \\
&\times&\left. \int dx \Big\{ g\left[1-\cos(qx)\right]\left[g_{\uparrow\uparrow}(x)+g_{\downarrow\downarrow}(x)\right]+\tilde{g}\left[1+\cos(qx)\right]g_{\uparrow\downarrow}(x) \Big\}\frac{d^2\delta(x)}{dx^2}\right] \;,
\label{Sqomega7}
\end{eqnarray}
where $\langle\Psi_0|T|\Psi_0\rangle$ is the expectation value on the ground state of the kinetic energy operator $T=-\left(\frac{\hbar^2}{2m}\sum_{i=1}^{N_a} \frac{\partial^2}{\partial x_i^2}+\frac{\hbar^2}{2m}\sum_{\alpha=1}^{N_b}\frac{\partial^2}{\partial x_\alpha^2}\right)$ and $g_{\sigma\sigma^\prime}(x)$ are the pair correlation functions for parallel and anti-parallel spins corresponding to the two components of the mixture. We notice that, as ${\bf q}\to 0$, the leading contribution to $m_3$ is of order $q^2$ and depends on the inter-species coupling constant $\tilde{g}$.

In the long-wavelength limit the fluctuation operator $\rho_{\bf q}^s$ acting on the ground state excites both the spin-wave (single-mode) state $|\Psi_{\text{sw}}\rangle$ with energy $\epsilon_s(q)\sim q$, as well as multi-mode states $|\Psi_{\text{mm}}\rangle$ with energy $\epsilon_{\text{mm}}\sim \text{const}$. From the definition (\ref{Sqomega2}) of the dynamic structure factor in terms of the matrix elements of the operator $\rho_{\bf q}^s$ and from the susceptibility sum rule (\ref{Sqomega3}) one finds that, in the long wavelength limit, the matrix element involving the spin-wave state should vanish as $|\langle\Psi_{\text{sw}}|\rho_{\bf q}^s|\Psi_0\rangle|^2\sim q$. On the contrary, the asymptotic behavior of $\langle\Psi_{\text{mm}}|\rho_{\bf q}^s|\Psi_0\rangle$ can be deduced from the $m_3$ sum rule (\ref{Sqomega7}) yielding: $|\langle\Psi_{\text{mm}}|\rho_{\bf q}^s|\Psi_0\rangle|^2\sim q^2$. As a consequence, at low momenta, the sum rules $m_{-1}$ and $m_0$ are both exhausted by the spin-wave state and multi-mode states contributions occur at higher order. Instead, spin-wave and multi-mode states contribute to $m_1$ at the same order in $q$. A useful remark is that the $m_3$ sum rule for the density fluctuation operator $\rho_{\bf q}=\sum_{i=1}^{N/2}e^{-iqx_i}+\sum_{\alpha=1}^{N/2}e^{-iqx_\alpha}$ vanishes as $q^4$ when ${\bf q}\to 0$. In this case also the $f$-sum rule $m_1$ is exhausted by the single-mode state (phonon mode) at long wavelengths~\cite{PinesNozieres66}.